\documentstyle[12pt]{article}
\begin{document}

\vskip -1.0cm
\rightline{\vbox{\hbox{CLNS 98/1575}}}

\vskip 0.4cm
\centerline {\Large \underline 
      {\bf An Update on CLEO's Study of B Meson Decays\footnotemark} }
\footnotetext{Lecture presented at the Quarks-98 Conference, May 18-23, 1998; 
Suzdal, Russia}

\vskip 0.2cm
\centerline {\large {\bf Bernard Gittelman} }
\vskip 0.1cm
\centerline {\bf August 1, 1998}
\vskip 0.4cm

\begin{abstract}
This is a discussion of the main branching fractions of B Meson decays
($b\rightarrow cW^-$).  A summary of what has been measured and what remains
unmeasured is presented.
\end{abstract}

\vskip 0.1cm
\noindent {\large {\bf Introduction }}
\vskip 0.2cm

The $\Upsilon$ family of mesons was discovered at Fermilab[1] in 1977.  At that
time, purely by coincidence the CESR storage ring had been designed and was
under construction.  Construction of the CLEO-I detector had also begun.  In
1978, the existence of the $\Upsilon(1S)$ and $\Upsilon(2S)$ resonances were 
confirmed at the DORIS storage ring of DESY[2], and from their measurement of
the $e^{+}e^{-} \rightarrow \Upsilon(1S)$ crossection, it became clear that the
$\Upsilon$ resonances are best described as $b\bar{b}$ quark pairs, not
$t\bar{t}$.  CESR began operation at the end of 1979. The CLEO (and CUSB)
Collaborations began data taking.  In 1980, CLEO (and CUSB) published
observation of the $\Upsilon(1S)$, $\Upsilon(2S)$, $\Upsilon(3S)$[3], and the
$\Upsilon(4S)$[4].  The $\Upsilon(4S)$, is a much wider resonance and is
believed to decay exclusively to $B\bar{B}$ mesons.  Over the next six to eight
years, many general properties of B meson decay were measured at DORIS and
CESR.  In 1988, CESR shut down and the CLEO-I detector was replaced by the
CLEO-II detector.  The new CLEO-II detector consists of a better resolution
tracking system and electromagnetic calorimeter.  Figure 1 is an illustration
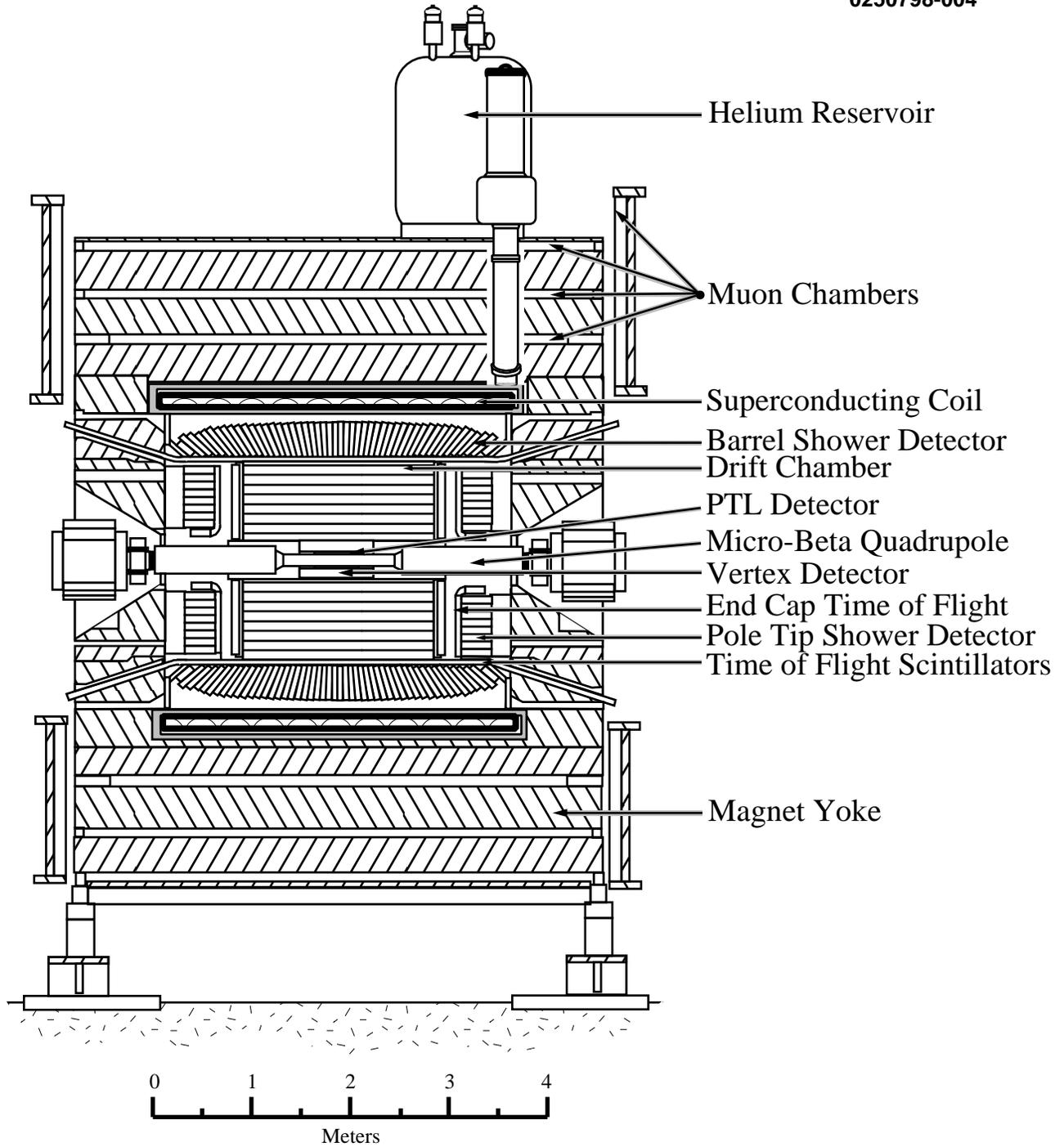
of the CLEO-II detector.  Our understanding of the B meson decay has continued
to improve with the CLEO-II detector, but today we only understand
approximately $50\%$ of the B decays.  This paper is a description of the
status of our understanding today compared with the Particle Data Group review
published in 1996.

\vskip 0.4cm
\noindent {\large {\bf Understanding of B Meson Decay in 1996 }}
\vskip 0.2cm

In the 1996 Review of Particle Physics[5], the Particle Data Group provided a
long detailed summary of what is known about the B meson branching fractions,
and life times.  Most of the information they reported on the B meson branching
fractions came from publications of the ARGUS Collaboration at DESY and
publications of CLEO at Cornell.  A summary of the ``largest'' B meson
branching fractions listed in the 1996 Review of Particle Physics is given in
Table 1.  In the Standard Model, the b quark decays under the weak interaction
to the c or u quark.  From early experiments, it was evident that the
$b\rightarrow u$ interaction is more than an order of magnitude weaker than the 
$b\rightarrow c$ (See for example the $b \rightarrow ul\nu$ and the $B
\rightarrow X_{c}l\nu$ branching fractions listed in Table 1.).  B meson decays
are usually described by the diagrams shown in figure 2 (Since the main decays
are to c quarks, the diagrams shown in figure 2 do not include the
$b\rightarrow u$ transitions.).  The first set of branching fractions in Table
1 are the semileptonic decay modes. These are the simplest decay modes to
describe.  They all proceed via the tree diagram. The listed branching
fractions, for example $Br(B \rightarrow Xl\nu) = (10.4 \pm 0.4)\%$, is
supposed to represent the branching fraction for $l=e$ or $l=\mu$.   The total
semileptonic branching fraction includes $B\rightarrow Xe\nu, B\rightarrow
X\mu\nu$, and $B\rightarrow X\tau\nu$. The branching fraction to electons and
muons are expected to be very close to identical.  However, the branching
fraction to $\tau$ leptons, which have never been measured, are expected to be
only $0.3$ as large, because of the large $\tau$ mass  (The factor 0.3 comes
from the phase space).  The sum over electron, muon and tau semileptonic decay 
is $2.3\times(10.4\pm 0.4) =  (23.9 \pm 0.9)\%$.

The second group of branching fractions in Table 1 are the 2 body decays of the
B meson in which the W boson materializes into a $u\bar{d}$, (i.e., $\pi^+$  or
$\rho^+$).  One notices that the $B^+$ branching fractions are 1.5 - 2 times
larger than the $B^o$'s.  The diagrams describing the $B^+\rightarrow
\pi^{+}D^{o}$ decay are the tree and the color supressed shown in figure 2. 
The $B^{o}\rightarrow \pi^{+}D^{-}$ decay proceeds through the tree and the
$W$ exchange diagram.  If the $W$ exchange amplitude is as large as the
color supressed amplitude, this would imply that color supressed amplitude is
in phase with the tree, and the $W$ exchange amplitude is out of phase.

The third group of branching fractions in Table 1 are the 2 body decays of the 
B meson into a $D^{(*)}D_{s}^{(*)}$.  These exclusive two body modes add up to 
$\approx (5.5 \pm 2.0)\%$.  The inclusive yield of $D_{s}$ coming from B decays
was listed as $(8.6\pm 2.2)\%$, which implies $64\%$ of the $B\rightarrow XD_s$
are two body.

\vspace{+0.2in}
\noindent Table 1, {\underline{\bf\large Summary of Major B Meson Branching 
	Fractions known in 1996}}
\vspace{-0.1in}
{ \bf
\begin{tabbing}
\hspace{1.1in} { ( Numbers from PDG Review[5] )} \\
\vspace{0.1in}
\hspace{1.0in} \=Decay Mode \qquad \= 
	\quad $B^+$ \qquad \qquad \= \quad $B^o$ 
        \qquad \qquad \= Admixture \\
 \>      \>$\quad(\%)$     \>$\quad(\%)$   \>$B^{+}$ and $B^{o}$ $(\%)$ \\
 \>$B \rightarrow \bar{D}l\nu$ \> $1.6\pm 0.7$ \> $1.9 \pm 0.5$ \\ 
 \>$B \rightarrow \bar{D}^{\ast}l\nu$ \> $5.3\pm 0.8$ \> $4.6 \pm 0.3$ \\ 
 \>$B \rightarrow \bar{D}^{\ast\ast}l\nu$ \>\quad \>\quad  \> $2.7\pm 0.7$ \\
 \>$B \rightarrow X_{c}l\nu$ \> $10.1\pm 2.3$ \> $10.3 \pm 1.0$ \> \\ 
 \>$B \rightarrow Xl\nu$ \> \quad   \> \quad   \> $10.4 \pm 0.4$  \\
 \>$b \rightarrow ul\nu$ \> \quad   \>  \quad   \> $\sim 0.18 \pm 0.05$ \\ 
	\> \\
 \>$B \rightarrow \bar{D}\rho^{+}$  \> $1.34 \pm 0.18$ \> $0.78 \pm 0.14$ \\
 \>$B \rightarrow \bar{D}^{\ast}\rho^{+}$\> $1.55\pm 0.31$ \> $0.73\pm 0.15$ \\
 \>$B \rightarrow \bar{D}\pi^{+}$  \> $0.53 \pm 0.05$ \> $0.30 \pm 0.04$ \\
 \>$B \rightarrow \bar{D}^{\ast}\pi^{+}$\> $0.52\pm 0.08$ \> $0.26\pm 0.04$ \\
 \>  \\
 \> $B \rightarrow \bar{D}D_{s}^{+}$  \>$1.7 \pm 0.6 $ \> $0.7 \pm 0.4 $ \\
 \> $B \rightarrow \bar{D}^{*}D_{s}^{+}$\>$0.1 \pm 0.07 $ \> $1.2 \pm 0.6 $ \\
 \> $B \rightarrow \bar{D}D_{s}^{*+}$  \>$1.2 \pm 1.0 $ \> $2.0 \pm 1.5 $ \\
 \> $B \rightarrow \bar{D}^{*}D_{s}^{*+}$\>$2.3 \pm 1.4 $ \> $1.9 \pm 1.2 $ \\
 \> $B \rightarrow XD^{+}_{s}$ \>\quad   \>\quad  \> $ 8.6 \pm 2.2$      \\
 \>  \\
 \>$B \rightarrow {D}^{-} X$   \>\quad   \>\quad   \> $24.2 \pm 3.3$ \\
 \>$B \rightarrow \bar{D}^{o} X$   \>\quad   \> \quad  \> $58.0 \pm 5.0$ \\
 \>  \\
 \>$B \rightarrow \Psi K$ \> $0.10\pm 0.014$ \> $0.075 \pm 0.021$   \\ 
 \>$B \rightarrow \Psi K^{+}\pi^{+}\pi^{-}$ \> $0.14\pm 0.06$ \\
 \>$B \rightarrow \Psi K^{+}\pi^{-}$ \>      \> $0.11 \pm 0.06$   \\ 
 \>$B \rightarrow \Psi K^{*}$ \> $0.17\pm 0.05$ \> $0.16 \pm 0.03$   \\ 
 \>$B \rightarrow \Psi X$   \>\quad   \> \quad  \> $1.14 \pm 0.06$ \\
	\> \\
 \>$B \rightarrow$ Charm-Baryons   \>\quad   \> \quad  \> $6.4 \pm 1.1$ \\
\end{tabbing} }
\vfil
\break

The inclusive decay of B mesons to $\bar{D}$ mesons in Table 1 add up to $(82.2
\pm 6.0)\%$.  These include the D mesons coming from semileptonic decay of the B
mesons.  Most of the rest of B meson decays are into Charmonium or 
Charm-Baryons.  Adding the $1.14\pm 0.06$ and the $6.4\pm 1.1$ to the
$82.2\pm 6.0\%$, one has accounted for $(89.7\pm 6.1)\%$ of B meson decays.

\vskip 0.4cm
\noindent {\large \bf{Inclusive B Meson Decay Branching Fractions} }
\vskip 0.2cm

During the past few years, CLEO has made many remeasurements of inclusive
branching fractions in B meson decay.  The most recent CLEO measurements on the
inclusive branching fractions are listed in Table 2.  These results are
generally consistent but a little more accurate than those shown in Table 1. 
The separate measurement of semileptonic decays of $B^+$ and $B^o$ mesons was
first made via full reconstruction of the $\bar{B}$ meson hadronic decay in the
event then examining how often there was a lepton coming from the remaining $B$
meson decay[6].  Although this method was very clear and simple, it was
statistically limited  by the small sample of fully reconstructed B (or
$\bar{B}$) mesons.  The results listed in Table 2 were obtained by partial
reconstruction of $\bar{B}^{o} \rightarrow D^{*+}l^{-}\nu, D^{*+}\rightarrow
\pi^{+}D^{o}$, detecting only the lepton and the $\pi^+$[7].  Having collected a
very large sample of partially reconstructed $B^o$s and measuring how often
there is also a lepton amongst the leftover tracks coming from the other B
meson decay, the semileptonic branching fraction of the $B^o$ was determined. 
To obtain the ${B}^{-} \rightarrow X^{o}l^{-}\nu$ branching fraction, the 
result of this analysis was then combined with a previous analysis[13] in which
the semileptonic branching fraction averaged over charged and neutral B mesons
was measured.  The result indicates the $B^o$ branching fraction is
$0.53\%$ larger than the $B^+$, which is indicates the $B^o$ lifetime is
$0.53\%$ longer than the $B^+$.  However, the uncertainty is large enough so
that the branching fractions could be equal.

The inclusive decays to D mesons in table 2 are the result of observing how
often it was possible to reconstruct a $D^o$ (via $D^{o}\rightarrow 
K^{-}\pi^{+}$) and how often it was possible to reconstruct $D^+$ (via
$D^{+}\rightarrow K^{-}\pi^{+}\pi^{+}$) in a set of events coming from
$\Upsilon(4S)$ decay.  The same analysis was used to measure the inclusive B
branching fraction to $D^{*+}$ and $D^{*o}$ by measuring how often it is
possible to reconstruct a $D^{*+}$ and how often a $D^{*o}$.  One should
understand the inclusive D branching fractions include D mesons coming from
$D^*$ decays.

Normally, the D mesons from $\bar{B}$ meson decay originate from the b quark
transition, $b\rightarrow cW^-$.  However, $\bar{D}$ mesons can come from
$\bar{B}$ meson decay by the $W^{-}\rightarrow \bar{c}s$.  Both the tree and
the color supressed diagrams (see figure 2) can lead to $\bar{D}$ from
$\bar{B}$ decays.  A measurement of how often the $\bar{B}$ meson decays to a D
meson and how often to a $\bar{D}$ meson has been made[9].  This was carried
out by using the fact that the $\Upsilon (4S)$ decays to $B\bar{B}$.  One of
the two B mesons is tagged via the sign of the lepton in its semileptonic
decay.  Only high momentum leptons ($p_{l} > 1.4$ GeV/c) are used, to eliminate
lepton from semileptonic D decay. The other B meson is assumed to have the
opposite flavor (Corrections are made for the $B^{o}\leftrightarrow
\bar{B}^{o}$ mixing of the neutral B mesons).  Having tagged the flavor of one
of the B mesons, the D meson coming from that same  B meson is known to have an
explicit flavor.  This D meson, which comes from the tagged B meson, is
identified via the angular correlation.  Since the lepton has a momentum
greater than 1.4 GeV/c, the D meson must be moving at an angle greater than
$90^o$ away from the lepton in order to conserve momentum.  The D mesons from
the other B meson are isotropic, so the flavor of the D mesons in the same
hemisphere as the lepton was studied.  From the numbers in Table 2, one sees
that $90\%$ of the D mesons from B decay come from the c quark that originates
from the $b\rightarrow cW^-$ transition and $10\%$ come from the W boson.

The $B\rightarrow X D_s$ inclusive branching fractions were measured by
studying the number of $D_s$ mesons and their momentum spectrum at the
$\Upsilon(4S)$  No attempt was made to tag the flavor of the B meson or its
charge.  From the $D_s$ momentum spectrum it was concluded that the $D_s$ is
almost always coming from the W boson.  To obtain the branching fraction,
$12.11\%$, it was assumed that the $B^+$ and $B^o$ branching fraction to $D_s$
are the same.  This value is significantly larger than the 1996 PDG value (see
Table 1.).

The most interesting new inclusive result is the measurement of the B decay to
$\eta X$[12].  The $17.6\%$ branching fraction is much larger than one might
expect.  Some of the $\eta$ mesons come from the D and $D_s$ mesons that are in
the B meson decay.  The contribution from feed down has been estimated from the
inclusive $B\rightarrow \bar{D^o}, D^{-}$, and $D_s$ branching fractions and
the inclusive $\eta$ branching fractions of each of these three charm mesons. 
The branching fraction estimates are ${\cal B}r(B\rightarrow
\bar{D}X\rightarrow \eta X') = (2.5\pm 0.5)\%$, and ${\cal B}r(B\rightarrow
D_{s}X\rightarrow \eta X') = (2.8\pm 0.6)\%$.  Subtracting these leads to the
conclusion $(12.3\pm 1.8)\%$  of B decays 

\vspace{+0.2in}
\noindent Table 2, {\underline
               {\bf\large Inclusive B Meson Branching Fractions from CLEO}}
{ \bf
\begin{tabbing}
\hspace{1.0in} \=\quad Decay Mode \qquad \qquad \quad \quad  \= 
	 Branching Fraction ($\%$)  \qquad \= Reference \\
   \> \\
   \>\underline{Semileptonic Decays} \\
   \>$B^{+} \rightarrow X^{o}l^{+}\nu$ \>$10.25\pm 0.57\pm 0.65$
		\>\qquad [7] \\ 
   \>${B}^{o} \rightarrow X^{-}l^{+}\nu$ \>$10.78\pm 0.60\pm 0.69$ 
		\>\qquad [7] \\ 
	\> \\
   \>\underline{Inclusive Decays to D mesons} \\
   \> Assuming an Equal Admixture of $B^{o}\bar{B}^{o}$ and $B^{+}{B}^{-}$  \\
   \>$B\rightarrow D^{o}X + \bar B\rightarrow D^{o}X$  \> 
		$63.6\pm 1.4\pm 2.6$ \> \qquad [8] \\
   \>$B\rightarrow D^{+}X + \bar B\rightarrow D^{+}X$
		 \> $23.5\pm 0.9\pm 2.4$ \> \qquad [8] \\
   \>	\\
   \>$B\rightarrow D^{\ast o}X + \bar B\rightarrow D^{\ast o}X$
		 \> $24.7\pm 1.2\pm 2.5$ \> \qquad [8] \\
   \>$B\rightarrow D^{\ast +}X + \bar B\rightarrow D^{\ast +}X$
		 \> $23.9\pm 1.5\pm 1.7$ \> \qquad [8] \\
	\> \\
   \>\underline{Inclusive Flavor Sensitive Decays to D mesons} \\
   \> (lv) for D meson from $b\rightarrow c$, 
             (uv) for D meson from $W^{-}\rightarrow \bar{c}s$ \\ 
   \>$\bar B\rightarrow D^{o}X + \bar B\rightarrow D^{+}X$ (lv)
        \>\quad $79.2\pm 3.9 $ \> \qquad [9] \\
   \>$\bar B\rightarrow \bar D^{o}X + \bar B\rightarrow D^{-}X$ (uv)
        \>\quad $\;\;7.9\pm 2.2 $ \> \qquad [9] \\
	\> \\
\>\underline{Inclusive B $\rightarrow D_{s}$ Decays} \\
   \> $B^{+}\rightarrow D_{s}^{+}X = B^{o}\rightarrow D_{s}^{+}X$
        \> $12.11\pm 0.39\pm 1.63 $ \> \qquad [10] \\
	\> \\
\>\underline{Inclusive B $\rightarrow$ Charmonium Decays} \\
   \>$ B\rightarrow \Psi X = \bar B\rightarrow \Psi X$
        \> $1.12\pm 0.04\pm 0.06 $ \> \qquad [11] \\
   \>  \\
   \>$ B\rightarrow \Psi X = \bar B\rightarrow \Psi X$ Direct
        \> $0.80\pm 0.08 $ \> \qquad [11] \\
   \>$ B\rightarrow \Psi' X = \bar B\rightarrow \Psi' X$
       \> $0.34\pm 0.04\pm 0.03 $ \> \qquad [11] \\
   \>$ B\rightarrow \chi_{c1}X = \bar B\rightarrow \chi_{c1}X$
       \> $0.40\pm 0.06\pm 0.04 $ \> \qquad [11] \\
   \>$ B\rightarrow \chi_{c2}X = \bar B\rightarrow \chi_{c2}X$
       \> $0.25\pm 0.10\pm 0.03 $ \> \qquad [11] \\
   \>  \\
   \>$ B\rightarrow$ Charmonium (total)
       \> $1.79\pm 0.15 $ \> from above \\
	\> \\
\>\underline{Inclusive B Decays to $\eta X$} \\
   \>$(B\rightarrow \eta X) $\> $17.6\pm 1.1\pm 1.2$ \> \qquad [12] \\
\end{tabbing} }

\vfil
\break

produce $\eta$ mesons directly.  $\eta$ mesons are often described as partly an
$s\bar{s}$ resonance, but this $12.3\%$  branching fraction is much larger than
the incusive  ${\cal B}r(B\rightarrow  \phi X) = (3.5\pm 0.7)\%$[5].

\vskip 0.5cm
\noindent {\large \bf{Exclusive B Meson Decay Branching Fractions} }
\vskip 0.2cm

The sum of all the exclusive branching fractions listed in Table 1 add up to
$(15.6\pm 2.1)\%$, ($16.6\%$ for $B^+$, $17.4\%$ for $B^o$).  In recent years,
many other 2 body final states have been searched for.  Most of these are
``rare B decays''[14].  The exclusive final states listed in Table 1 have been
remeasured.  The new CLEO results for these and a few others are given in Table
3.  There have been several new measurements of semileptonic decays.  For the
$b\rightarrow c$ final states, there have been no serious changes.  The
$D^{o}_{1}l\nu$ has been measured.  There have also been measurements of the
$b\rightarrow u$ final states, $\pi l\nu$ and $\rho l\nu$.  The branching 
fractions of the exclusive $b\rightarrow cl\nu$ final states that have been
observed add up to $7.0\%$ which is only $\sim 67\%$ of the inclusive
branching fraction.  The observed $b\rightarrow u$ final states add up to
$0.043\%$ compared with the inclusive rate of $0.18\%$ from Table 1,
i.e., only $\sim 25\%$ of the $b\rightarrow ul\nu$ is understood.

\vspace{+0.2in}
\noindent Table 3, {\underline{\bf\large Exclusive B meson Decay Branching 
           Fractions from CLEO}}
{ \bf
\begin{tabbing}
\hspace{0.75in} \=\quad Decay Mode \quad \quad  \= 
	\quad $B^+$ \qquad \qquad \qquad \qquad \= \quad $B^o$ 
        \qquad \qquad \qquad \= Reference \\
        \>      \>$\quad(\%)$     \>$\quad(\%)$   \> \\
   \>\underline{Semileptonic Decays} \\
   \>$B \rightarrow Dl\nu$ \> $1.94\pm 0.15\pm 0.34$ \> $1.87\pm 0.15\pm 0.32$ 
            \> \qquad [15]\\ 
   \>$B \rightarrow D^{\ast}l\nu$ \> $5.13\pm 0.54\pm 0.64$ \> $4.49\pm 0.32
 		\pm 0.39$ \> \qquad [16] \\ 
   \>$B \rightarrow D^{o}_{1}l\nu$ \>$0.56\pm 0.13\pm 0.09$ \>\quad ? 
		\> \qquad [17] \\
   \> \\
   \>$B \rightarrow \pi l\nu$ \> \>$0.018\pm 0.004\pm 0.004$
	\> \qquad [18]\\ 
   \>$B \rightarrow \rho l\nu$ \> \>$0.025\pm 0.004\pm 0.007$ 
	\> \qquad [18] \\ 
	\> \\
   \>\underline{Exclusive Two Body Charm Decays} \\
    \> $B \rightarrow D\rho$  \> $1.35 \pm 0.12\pm 0.15$ 
	  \> $0.81 \pm 0.11\pm 0.18$ \> \qquad [19] \\
    \> $B \rightarrow D^{\ast}\rho$  \>$ 1.68\pm 0.21 \pm 0.28$
	     \> $0.74\pm 0.10\pm 0.14$ \> \qquad [19] \\
    \> $B \rightarrow D\pi$  \> $0.55\pm 0.04\pm 0.05$ 
	  \> $0.29\pm 0.04\pm 0.06$ \> \qquad [19] \\
    \> $B \rightarrow D^{\ast}\pi$  \> $0.434\pm 0.033\pm 0.038$
  	     \> $0.281\pm 0.011\pm 0.022$ \> \qquad [20]\\
   \>  \\
   \>$B \rightarrow DD_s^{+}$ \> $1.26\pm 0.22\pm 0.29$ 
            \> $0.87\pm 0.24\pm 0.22$ \> \qquad [21]\\ 
   \>$B \rightarrow DD_s^{\ast +}$ \> $0.87\pm 0.27\pm 0.20$ 
            \> $1.00\pm 0.35\pm 0.25$ \> \qquad [21]\\ 
   \>$B \rightarrow D^{\ast}D_s^{+}$ \> $1.40\pm 0.43\pm 0.39$ 
            \> $0.93\pm 0.23\pm 0.19$ \> \qquad [21]\\ 
   \>$B \rightarrow D^{\ast}D_s^{\ast +}$ \> $3.10\pm 0.88\pm 0.75$ 
            \> $2.03\pm 0.50\pm 0.43$ \> \qquad [21]\\ 
   \> \\
	\> $B \rightarrow D^{-}D^{+}$  \> \> \quad$< 0.12$ 
		\> \qquad [22]\\
	\> $B \rightarrow D^{\ast\pm}D^{\mp}$  \> \> \quad$< 0.18$ 
		\> \qquad [22] \\
	\> $B \rightarrow D^{\ast -}D^{\ast +}$  \> \> \quad$< 0.22$ 
		\> \quad [22,26]\\
	\> \\
   \>\underline{Exclusive B $\rightarrow$ Charmonium Decays} \\
   \>$B \rightarrow \Psi K$ \> $0.102\pm 0.008\pm 0.007$
		\> $0.085\pm 0.013\pm 0.006$\> \qquad [23] \\ 
   \>$B \rightarrow \Psi K^{\ast}$ \> $0.141\pm 0.023\pm 0.024$
		\>$0.132\pm 0.017\pm 0.017$ \> \qquad [23] \\ 
   \>$B \rightarrow \Psi \pi$ \> $0.0056\pm 0.0027$
		\> $< 0.0058$  \> \qquad [24] \\ 
   \>$B \rightarrow \Psi \rho$ \> $< 0.077$
		\>$ < 0.025$ \> \qquad [24] \\ 
   \>$B \rightarrow \Psi \omega$ \>
		\>$ < 0.027$ \> \qquad [24] \\ 

\end{tabbing} }


The exclusive two body charm decays in which the W boson materializes as a
$\pi, \rho$ or $D_s$ have been remeasured.  Most of the results agree with the
earlier measurements and are more accurate.  The remeasurements on
$B\rightarrow \Psi K^{(*)}$ have similar results.  

\vskip 1.3cm

\noindent {\large\bf
      {Summary- What Fraction of B Meson Decays are ``Understood''}}
\vskip 0.2cm

Theoretically, the Semileptonic decays of B mesons are the easiest to
understand, because there is only one hadronic vertex.  Originally[25], the 
semileptonic branching fraction was estimated to be approximately $16\%$. 
After independent measurements at DORIS, CESR, and LEP, we find the branching
fraction is significantly smaller.  The generic average of the CLEO
measurements gives the inclusive branching fraction,
$$ {\cal B}r(\bar{B} \rightarrow Xl\nu) = (10.5 \pm 0.89)\%. $$

\noindent Although using QCD corrections, theorists have recently suggested the
B semileptonic branching fraction could be as low as $12\%$, the CLEO result is
a theoretical challenge.  Furthermore, only a fraction of the of the hadronic
content of the semileptonic decays have been measured.  Summing over the
measured charm hadronic states, one has 
$$ {\cal B}r(\bar{B} \rightarrow (D + D^{*} + D^{**})l\nu) =  
		(7.00 \pm 0.85)\% $$

\noindent Therefore, we understand $(7.00 \pm 0.85)/(10.5 \pm 0.89) = 
(67.0 \pm 13.7)\%$ of semileptonic branching fractions.

\vskip 0.4cm

From the CLEO measurement of the flavor dependence of the decay of B mesons to
D mesons, we know the ${\cal B}r(\bar{B}\rightarrow DX) = (79.2\pm 3.9)\%$
(i.e., from $ b\rightarrow c$ at the lower vertex).  Assuming the exclusive 
semileptonic branching fractions to $\tau$'s are 0.3 times the branching
fractions to electrons and/or muons, and adding the exclusive semileptonic
decays to e's, $\mu$'s, and $\tau$'s, we have for the total exclusive 
semileptonic branching fraction
$$ {\cal B}r(\bar{B} \rightarrow D^{(*,**)}(e + \mu + \tau)\nu) = 
    (2.3\times (7.00 \pm 0.85) = (16.1 \pm 2.0)\% $$
\noindent The exclusive two body decays resulting in a D meson add up to 
$$ {\cal B}r(\bar{B} \rightarrow D\pi + D^{*}\pi + D\rho + D^{*}\rho +
       DD_{s} + D^{*}D_{s} + DD_{s}^{*} + D^{*}D_{s}^{*}) = (8.8\pm 1.2)\%$$ 
\noindent (This was calculated from the numbers in Table 3, averaging over
$B^+$ and $B^o$.)  Putting the above numbers together, 
the inclusive $B\rightarrow DX$ decays are ``understood'' at the level,
$$ {\cal B}r(\bar{B} \rightarrow D^{(*)}( l\nu + \pi + \rho + D_{s})) =
      (16.1 \pm 2.0) + (5.73\pm 1.17) + (3.07\pm 0.35) = (24.9\pm 1.8)\%$$
This is $(24.9\pm 1.8)/(79.2\pm 3.9) = 0.31\pm 0.03$ of the inclusive D mesons
coming from the $b\rightarrow c$ vertex.

The average of exclusive final states with charmonium that have been measured
add up to $\sim (0.23\pm 0.03)\%$.  The exclusive baryon final states have not
yet been fully reconstructed.  Therefore, we have an exclusive understanding of 
$(24.9\pm 1.8 + 0.23\pm 0.03) = (25.1\pm 1.8)\%$ of the B decays.

\vfil
\break

In terms of inclusive branching fractions, in the $\bar{B}\rightarrow DX$ 
decays, the X represents $l\nu X'$, $D_{s}X''$, and $X'''$.  Here $X'$ includes
the $\pi$'s coming from $D^*$ or $D^{**}$ which we know something about, but
$X''$ and $X'''$ are not known.  From the value of the branching fraction 
$\bar{B}\rightarrow DX = (79.2\pm 3.9)\%$, we calculate,  \begin{tabbing}
\hspace{1.0in}${\cal B}r(\bar{B}\rightarrow DX''')$\=  $= \bar{B}\rightarrow
D(X - l\nu X' - D_{s}X'')$ \\  \>$= (79.2\pm 3.9) - ( 2.3\times(10.5\pm 0.89))
- (12.1\pm 1.7)$ \\ \>$= (42.9 \pm 4.7)\%$ \\ \end{tabbing} \noindent This
means we know nothing about $(42.9/79.2) = 54\%$ of the  inclusive
$B\rightarrow DX$ decays.

\vskip 0.2cm
\noindent
The total inclusive branching fractions, (excluding the ``rare'' decays) add up
to 
\begin{tabbing}
\qquad \=${\cal B}r(B\rightarrow DX + charmoniumX' + baryonsX'') $\=
   $= (79.2\pm 3.9) + (1.8\pm 0.2) + (6.4\pm 1.1)$ \\
       \>     \> $= (87.4 \pm 4.1)\%$\\
\end{tabbing}

CLEO has recently made many measurements of rare decays (and upper limits of
rare decays)[14].  However, the total sum of the measured modes and the upper
limits on rare B decay  branching fractions is less than $2\times 10^{-3}$, so
there is no need to includes these at this time.  The conclusion is one must
invent non-rare decay modes and look for them.

\vskip 0.4cm
\noindent {\large {\bf Acknowledgements }}
\vskip 0.2cm
I would like to thank Giancarlo Moneti and Ed Thorndike of the CLEO
Collaboartion for useful comments, corrections, and discussions related to this
paper.  The hospitality of the organizers of the Quarks-98 Seminar in Suzdal
made this visit to Russia, my first in ten years, warm and interesting.

\vfil
\break

\centerline {\large \underline 
      {\bf References} }

\newcounter{nmbr}
\begin{list}
{\arabic{nmbr}.}{\usecounter{nmbr} \setlength{\rightmargin}{\leftmargin}}
\item Observation of a new group of resonances at 10 GeV decaying to muons pairs
 \newline S.W. Herb et al., Phys. Rev. Lett. $\underline{39}, 252(1977).$      
\item Confirmation of the existence of the $\Upsilon(1S)$ and $\Upsilon(2S)$
  \newline C. Berger et al., Phys. Lett. $\underline{76B}, 243(1978).$          
\newline C.W. Darden et al., Phys. Lett. $\underline{78B}, 246(1978).$        
\newline J.K. Bienlein et al., Phys. Lett. $\underline{78B}, 360(1978).$      
\newline C.W. Darden et al., Phys. Lett. $\underline{80B}, 419(1979).$        
\item  Observation of Three Upsilon States
 \newline D. Andrews  et al., Phys. Rev. Lett. $\underline{44}, 1108(1980).$   
\newline                                                                   
  T. Bohringer et al., Phys. Rev. Lett. $\underline{44}, 1111(1980).$ 
\item Observation of a Fourth Upsilon State in $e^+e^-$ Annihilations
 \newline D. Andrews et al., Phys. Rev. Lett. $\underline{45}, 219(1980).$     
\newline 
  G. Finnocchiaro et al., Phys. Rev. Lett. $\underline{45}, 222(1980).$ 
\item Particle Data Group Review, Phys.Rev.D $\underline{54}, 1(1996)$     
\item Semileptonic Branching Fractions of Charged and Neutral $B$ Mesons
 \newline M.Athanas et al., Phys. Rev. Lett. $\underline{73}, 3503(1994)$,     
 \newline \qquad
           (erratum, Phys. Rev. Lett. $\underline{74}, 3090(1995)$) 
\item Study of the $B^0$ Semileptonic Decay Spectrum at the $\Upsilon(4S)$
       Resonance
 \newline M. Artuso et al., Phys. Lett. B $\underline{399}, 321(1997).$        
\item The Inclusive Decays $B \to D X$ and $B \to D^* X$
 \newline L. Gibbons et al., Phys. Rev. D $\underline{56}, 3783(1997).$        
\item Flavor-Specific Inclusive $B$ Decays to Charm
 \newline T.E. Coan et al., Phys. Rev. Lett. $\underline{80}, 1150(1998).$     
\item Measurements of $B \to D^+_s X$ Decays
 \newline D. Gibaut et al., Phys. Rev. D $\underline{53}, 4734(1996).$         
\item Inclusive Decays of $B$ Mesons to Charmonium
 \newline R. Balest et al., Phys. Rev. D $\underline{52}, 2661(1995).$         
\item The Inclusive Decay $B \to \eta X$
 \newline Y. Kubota et al., Phys. Rev. D $\underline{53}, 6033(1996).$         
\item Measurement of the $B$ Semileptonic Branching Fraction with Lepton Tags
 \newline B. Barish et al., Phys. Rev. Lett. $\underline{76}, 1570(1996).$     
\vfil
\break
\item Search for Exclusive Charmless Hadronic $B$ Decays
 \newline D.M.Asner et al., Phys. Rev. D $\underline{53}, 1039(1996).$         
 \newline Evidence for Penguin-Diagram Decays: First Observation of $...$
 \newline R. Ammar et al., Phys. Rev. Lett. $\underline{71}, 674(1993).$
 \newline Observation of $B^+ \to \omega K^+$ and Search for Related
	$B$ Decay Modes
 \newline T. Bergfeld et al., Phys. Rev. Lett. $\underline{81}, 272(1998).$
\item Measurement of the $\bar{B} \to D \ell \bar{\nu}$ Partial Width and Form
	Factor Parameters
 \newline M. Athanas et al., Phys. Rev. Lett. $\underline{79}, 2208(1997).$    
\item Measurement of the $\bar{B} \to D^* \ell \bar{\nu}$ Branching Fractions
	and $|V_{cb}|$
 \newline B. Barish et al., Phys. Rev. D $\underline{51}, 1014(1995).$         
\item Investigation of Semileptonic $B$ Meson Decay to P-Wave Charm Mesons
 \newline A. Anastassov et al., Phys. Rev. Lett. $\underline{80}, 4127(1998)$. 
\item First Measurement of the 
   $B \to \pi \ell \nu$ and $B \to \rho(\omega) \ell \nu$ Branching Fractions
 \newline J. Alexander et al., Phys. Rev. Lett. $\underline{77}, 5000(1996).$  
\item Exclusive Hadronic $B$ Decays to Charm and Charmonium Final States
 \newline M.S. Alam et al., Phys. Rev. D $\underline{50}, 43(1994).$           
\item New Measurement of $B \to D^* \pi$ Branching Fractions
 \newline G.Brandenburg et al., Phys. Rev. Lett. $\underline{80}, 2762(1998).$ 
\item Measurements of $B \to D^+_s X$ Decays
 \newline D. Gibaut et al., Phys. Rev. D $\underline{53}, 4734(1996).$         
\item Search for the Decays $B^0 \to D^{(*)+} D^{(*)-}$
 \newline D.M. Asner et al., Phys. Rev. Lett. $\underline{79}, 799(1997).$     
\item Measurement of the Decay Amplitudes and Branching Fractions 
	of $B \to J/\psi K^*$ and $B \to J/\psi K$ Decays
 \newline C.P. Jessop et al., Phys. Rev. Lett. $\underline{79}, 4533(1997).$   
\item Study of $B \to \psi \rho$
 \newline M. Bishai et al., Phys. Lett. B $\underline{369}, 186(1996).$        
\item B Meson Decay, B. Gittelman, S. Stone 
 \newline High Energy Electron-Positron Physics, World Scientific 
	(Singapore), (1988)
\item Recently there has been a measurement of the branching fraction,
      ${\cal B}r(B\rightarrow D^{*-}D^{*+}) = (7.8^{+5.4}_{-3.8} \pm 1.5)\times
      10^{-4}$.  This has been submitted to the Van Couver conference, 
      CLEO CONF 98-07; ICHEP98 849.
\end{list}

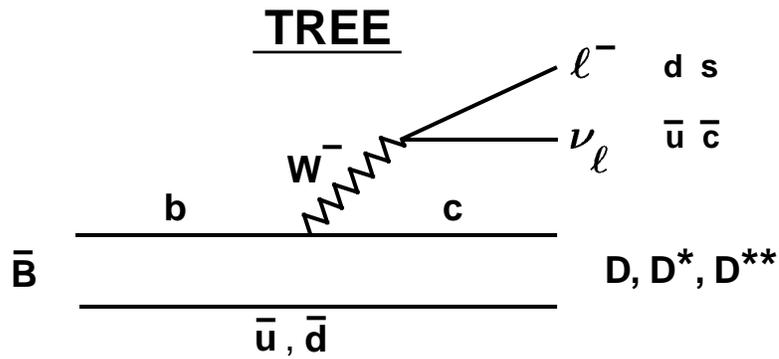
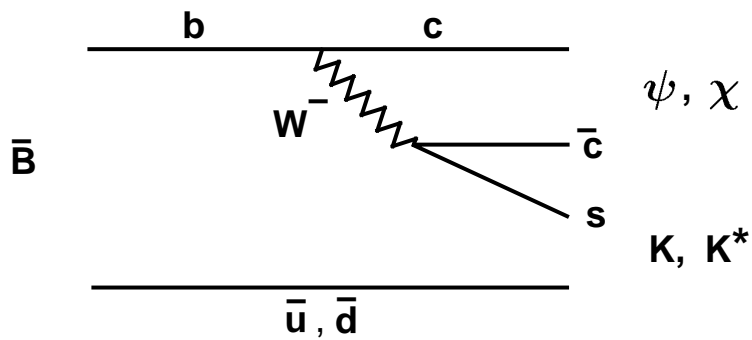
\end{document}